\documentclass[journal]{IEEEtran}
\usepackage{cite,url,subfigure,epsfig,graphicx}
\usepackage{tipa}
\usepackage{makecell}
\usepackage{multirow}
\usepackage{bbm}
\usepackage{mathrsfs}
\usepackage{array}
\usepackage{amsfonts}
\usepackage{cite,url,subfigure,epsfig,graphicx}
\usepackage{amssymb,amsmath}
\usepackage{indentfirst}
\usepackage{pifont} %number with cycle
\usepackage{algorithmic}
\usepackage{algorithm}
\usepackage{cases}
\IEEEoverridecommandlockouts
\makeatletter
\usepackage{times,verbatim,amsfonts,amsmath,color}
\begin{document}

\title{Combating Advanced Persistent Threats: Challenges and Solutions}

\author{
\IEEEauthorblockN{Yuntao~Wang\IEEEauthorrefmark{2}, Han~Liu\IEEEauthorrefmark{2}, Zhendong~Li\IEEEauthorrefmark{3}, Zhou~Su\IEEEauthorrefmark{2}\IEEEauthorrefmark{1}, and Jiliang~Li\IEEEauthorrefmark{2}}\\
\IEEEauthorblockA{
\IEEEauthorrefmark{2}School of Cyber Science and Engineering, Xi'an Jiaotong University, China\\
\IEEEauthorrefmark{3}School of Information and Communication Engineering, Xi'an Jiaotong University, China
}
\thanks{This work has been accepted by IEEE NETWORK in April 2024.}}

\maketitle
\begin{abstract}
The rise of advanced persistent threats (APTs) has marked a significant cybersecurity challenge, characterized by sophisticated orchestration, stealthy execution, extended persistence, and targeting valuable assets across diverse sectors. Provenance graph-based kernel-level auditing has emerged as a promising approach to enhance visibility and traceability within intricate network environments. However, it still faces challenges including reconstructing complex lateral attack chains, detecting dynamic evasion behaviors, and defending smart adversarial subgraphs. To bridge the research gap, this paper proposes an efficient and robust APT defense scheme leveraging provenance graphs, including a network-level distributed audit model for cost-effective lateral attack reconstruction, a trust-oriented APT evasion behavior detection strategy, and a hidden Markov model based adversarial subgraph defense approach. Through prototype implementation and extensive experiments, we validate the effectiveness of our system. Lastly, crucial open research directions are outlined in this emerging field.
\end{abstract}

\IEEEpeerreviewmaketitle
%\textcolor{blue}{ }
\section{Introduction}
Advanced persistent threats (APT) \cite{2} has emerged as a significant cybersecurity threat characterized by highly organized and well-funded attackers, stealthy and evasive execution, long-term persistence, and precise targeting of high-value assets. APT attacks can have devastating consequences across various sectors, including government, critical infrastructures, corporations, and individuals. The objectives of APT attacks often encompass espionage, theft of sensitive information, intellectual property, financial gain, and disruption of critical information infrastructures. Based on statistics from 360 Security\footnotemark[1], APT (e.g., Stuxnet, Gauss, Flame, and Duqu) constitutes nearly 60\% of cyberattacks targeting governments, transnational corporations, and critical infrastructures over the last two years.
\footnotetext[1]{\url{https://sc.360.net/}}
%\cite{360report}repeated attempts,
A typical APT attack lifecycle comprises the following steps \cite{3}.
\begin{itemize}
    \item \textit{Initial Compromise:} APT attackers establish their foothold through tactics such as spear-phishing emails, social engineering, watering hole attacks, or exploiting software vulnerabilities. This initial compromise serves as a starting point for {attackers} to infiltrate the target network.
    \item \textit{Lateral Movements:} APTs are typically orchestrated by a team of sophisticated hackers, working in a coordinated fashion.
    Once inside the network, APT attackers can employ diverse techniques to move laterally across systems. This involves escalating privileges, exploiting weak credentials, and leveraging known vulnerabilities to gain access to vital assets.
    \item \textit{Persistence:} APT attackers ensure their continued access by implementing {mechanisms} such as backdoors, Trojans, or remote access tools. These mechanisms enable them to maintain control and re-enter compromised systems even after being detected.
    \item \textit{Data Exfiltration:} APT attackers meticulously identify and exfiltrate sensitive data over an extended period. This step necessitates a deep understanding of the victim's data landscape and careful evasion of security measures.
\end{itemize}

To combat the complex and evolving nature of APT attacks, provenance graph-based kernel-level auditing \cite{2,3} offers a promising approach {with enhanced} visibility, traceability, and detection capabilities within intricate and dynamic network environments. It involves real-time capturing and analysis of intricate system interactions, encompassing network communications, process interactions, and file operations. By constructing causal relationship graphs of these entities, the provenance graph provides an all-encompassing depiction of system behaviors, yielding the following advantages \cite{3}:
\begin{itemize}
    \item \textit{Traceability:} The provenance graph facilitates the tracing of actions and interactions within a system, streamlining the identification of suspicious or malicious behaviors.

 \item \textit{Real-time Visibility:} Through the real-time capture of low-level system activities, the provenance graph delivers a dynamic comprehension of ongoing processes and potential threats.

\item \textit{Covert Behavior Detection:} The provenance graph aids in the revelation of concealed APT activities that may elude traditional detection mechanisms.

\item \textit{Attack Reconstruction:} Leveraging the provenance graph, security analysts can reconstruct the sequence of point-of-interest (PoI) events leading to an attack, thus assisting in post-incident analysis and response.
\end{itemize}

\begin{figure*}[!t]
\setlength{\abovecaptionskip}{-0.0cm}
\centering
  \fbox{\includegraphics[width=17cm]{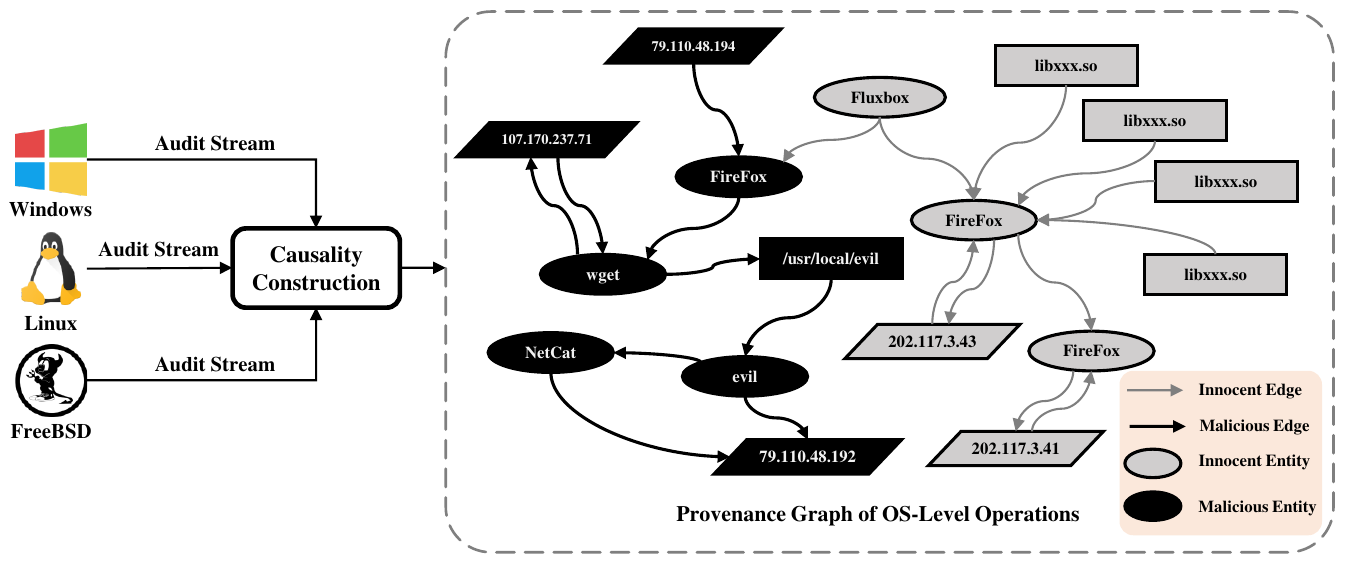}}
  \caption{An Overview of Provenance Graph-Based APT Audit Approach.}\label{fig:1}\vspace{-2mm}
\end{figure*}

However, the provenance graph-based kernel-level APT audit technology encounters the following new challenges.

\begin{itemize}
    \item \textit{Reconstruction of Lateral Attack Chains:}
    Adversaries can breach system boundaries through highly covert attacks, such as leveraging zero-day vulnerabilities or backdoors. They exploit lateral movements and domain controller hijacking in the target intranet to establish specific hop chains, triggering security alerts such as data exfiltration, password cracking, and shellcode payloads. As such, it is challenging for traditional host-based provenance intrusion detection systems (Prov-HIDS) to fully reconstruct APT attack patterns \cite{3}. Additionally, the host-level provenance graph usually contains millions of data entities \cite{6}, leading to dependency explosion problems during provenance graph audits, thereby impacting the availability of APT provenance services.

\item \textit{Identification of APT Evasion Behaviors:}
Recent studies \cite{5,6} highlight that real APT attacks often utilize strategic tactics, such as integrating numerous unrelated inter-process communication (IPC) sequences into attack primitives to evade provenance graph-based APT audits (refer to as APT evasion behaviors). Moreover, the varied functional deployment of network devices (e.g., switches, DNS servers, and domain controllers) complicates {the design of} compatible network-level provenance graph {methods}, amplifying the intricacy of identifying APT evasion behaviors.

\item \textit{Adversarial Subgraph Detection:}
Prov-HIDS systems generally rely on subgraph matching \cite{7,8} and cyber threat intelligence (CTI) to simulate APT behaviors for matching and auditing. Nevertheless, provenance graphs are vulnerable to adversarial attacks. For example, adversaries can craft adversarial sub-provenance graphs \cite{9} that avoid disrupting attack primitives, thereby evading detection through matching. Consequently, it diminishes the effectiveness of provenance graph auditing.
\end{itemize}

Hence, it is urgent to design a robust and efficient APT detection scheme based on provenance graphs, with the ability to reconstruct APT lateral movements, detect APT evasion behaviors, and uncover adversarial subgraphs.

As an effort to address the above challenges, this paper proposes a novel provenance graph based APT defense approach with low complexity and high robustness. Specifically, we present a general architecture of network-layer provenance graph-based APT audit. Then, under this architecture, we devise three components: (i) network-level distributed provenance graph audit model for cost-effective lateral attack chain reconstruction, (ii) trust-oriented dynamic APT evasion behavior detection strategy for improved availability of APT defense services, and (iii) hidden Markov model (HMM)-based adversarial subgraph detection strategy for enhanced robustness of APT defense services. Finally, we implement a real prototype and carry out extensive experiments to validate the feasibility and effectiveness of our proposed system.

The remainder of this paper is organized as follows. Section~II shows the working principle and key challenges of provenance graph-based APT audit. Section~III presents the proposed solutions under the provenance graph-based APT audit architecture. Section~IV demonstrates the prototype implementation and experimental evaluation. Section~V outlines future research directions, and Section~VI concludes this work with conclusions.

\section{Working Principle and Challenges of Provenance Graph-Based APT Audit}

\begin{table*}[!t]\scriptsize
\setlength{\abovecaptionskip}{-0.0cm}
\begin{center}
\caption{A comparison of our work with the state-of-the-arts in provenance graph-based APT defense (PG: provenance graph)}\label{table1}
\begin{tabular}{|c|c|c|c|c|c|c|c|}
\hline
\textbf{Ref}  & \textbf{Method} & \textbf{Advantages} & \textbf{Limitations} & \textbf{\begin{tabular}[c]{@{}c@{}}Priori \\ knowledge\end{tabular}} & \textbf{\begin{tabular}[c]{@{}c@{}}Network-level \\ PG audit\end{tabular}} & \textbf{\begin{tabular}[c]{@{}c@{}}APT evasion \\ attack\end{tabular}} & \textbf{\begin{tabular}[c]{@{}c@{}}Adversarial \\ attack\end{tabular}} \\ \hline

{StreamSpot}{[12]}        &      \begin{tabular}[c]{@{}c@{}}Sketch-based\\  Prov\end{tabular}           &    \begin{tabular}[c]{@{}c@{}}Dynamically \\ Maintainable\end{tabular}          &   \begin{tabular}[c]{@{}c@{}}Work for\\Small-scale Graph\end{tabular}            &                                                         \ding{56}              & \ding{56}                                                                                       & \ding{56}                                                                     & \ding{56}                                                                     \\ \hline
{UNICORN}{[6]}        &      \begin{tabular}[c]{@{}c@{}}Sketch-based\\  Prov\end{tabular}          &    \begin{tabular}[c]{@{}c@{}}Slow-acting\\  Attack Defense\end{tabular}           &      \begin{tabular}[c]{@{}c@{}}High-false\\ Alarm Rate\end{tabular}         &                                                         \ding{56}              & \ding{56}                                                                                       & \ding{52}                                                                     & \ding{56}                                                                     \\ \hline
{ProvDetector}{[7]}        &     \begin{tabular}[c]{@{}c@{}}Stealthy Malware\\  Detection\end{tabular}          &      \begin{tabular}[c]{@{}c@{}}Hidden Attack\\  Detection\end{tabular}        &    \begin{tabular}[c]{@{}c@{}}Only Support\\ Offline Detection\end{tabular}          &                                                         \ding{52}              & \ding{56}                                                                                       & \ding{56}                                                                     & \ding{56}                                                                     \\ \hline
{SLEUTH}{[9]}        &      \begin{tabular}[c]{@{}c@{}}Dependency Graph\\  Abstraction\end{tabular}           &     \begin{tabular}[c]{@{}c@{}}Attack Scenario\\  Reconstruction\end{tabular}           &       \begin{tabular}[c]{@{}c@{}}Prone to\\ 0day Threats\end{tabular}       &                                                         \ding{52}              & \ding{56}                                                                                       & \ding{56}                                                                     & \ding{56}                                                                     \\ \hline
{NODOZE}{[10]}        &      \begin{tabular}[c]{@{}c@{}}Frequency \\  Dependency Prov\end{tabular}            &       \begin{tabular}[c]{@{}c@{}} Entities Diffusion\\ Analysis\end{tabular}         &        \begin{tabular}[c]{@{}c@{}} Covert Attack\\ Ineffectiveness\end{tabular}        &                                                         \ding{56}              & \ding{56}                                                                                       & \ding{56}                                                                     & \ding{56}                                                                     \\ \hline
{Poirot}{[11]}        &     \begin{tabular}[c]{@{}c@{}}CTI Graph\\Alignment\end{tabular}            &   \begin{tabular}[c]{@{}c@{}}CTI Reconstruction\\ Graph\end{tabular}             &       \begin{tabular}[c]{@{}c@{}}Need Excessive\\ Prior Knowledge\end{tabular}         &                                                         \ding{52}              & \ding{56}                                                                                       & \ding{56}                                                                     & \ding{56}                                                                     \\ \hline
{ HOLMES}{[12]}        &     \begin{tabular}[c]{@{}c@{}}High-level\\  Scenario Graph\end{tabular}           &        \begin{tabular}[c]{@{}c@{}}Advanced Semantic\\ Mapping\end{tabular}        &        \begin{tabular}[c]{@{}c@{}}Requisite for\\ Extensive Expertise\end{tabular}        &                                                         \ding{52}              & \ding{56}                                             & \ding{56}                                                                     & \ding{56}                                      \\ \hline
%{Extrator }{[13]}        &     \begin{tabular}[c]{@{}c@{}}Natural Language\\  Processing\end{tabular}            &     \begin{tabular}[c]{@{}c@{}}Subgraph \\ Matching\end{tabular}           &      \begin{tabular}[c]{@{}c@{}}Requisite for\\ Extensive Expertise\end{tabular}           &                                                         \ding{52}              & \ding{56}                                          & \ding{56}                                                                     & \ding{56}                                                                     \\ \hline
{ATLAS }{[1]}        &     \begin{tabular}[c]{@{}c@{}}End-to-end\\  Attack Story\end{tabular}            &          \begin{tabular}[c]{@{}c@{}}End-to-end \\ Traceability\end{tabular}      &     \begin{tabular}[c]{@{}c@{}} Covert Attack\\ Ineffectiveness\end{tabular}          &                                                         \ding{52}              & \ding{56}                                                                                       & \ding{56}                                                                     & \ding{56}                                                                     \\ \hline
{DEPIMPACT}{[14]}        &      \begin{tabular}[c]{@{}c@{}}Dependency\\ Graph Weight\end{tabular}            &          \begin{tabular}[c]{@{}c@{}} Causality  Graph \\Reconstruction\end{tabular}      &        \begin{tabular}[c]{@{}c@{}} Prone to Poisoning \\Attack\end{tabular}        &                                                         \ding{56}              & \ding{56}                                                                                       & \ding{56}                                                                     & \ding{56}                                                                     \\ \hline
{PROVNINJA}{[15]}        &          \begin{tabular}[c]{@{}c@{}} Process Gadget\\ Chains\end{tabular}        &           \begin{tabular}[c]{@{}c@{}} Adversarial Attack \\Achievement\end{tabular}     &      \begin{tabular}[c]{@{}c@{}} Lack of  \\Defensive Measure\end{tabular}          &                                                              \ding{56}                 & \ding{56}                                                                                       & \ding{56}                                                                     & \ding{52}                                                                     \\ \hline
\textbf{Ours} &         \begin{tabular}[c]{@{}c@{}}CPA+LDA,\\Trust model,\\HMM \end{tabular}          &       \begin{tabular}[c]{@{}c@{}}Network-level Audit,\\APT Evasion-resist,\\ Adversarial Robust \end{tabular}            &             \begin{tabular}[c]{@{}c@{}}Only Prototype,\\Lack of Large-scale\\Actual Deployment \end{tabular} &                  \textbf{Partly}                                & \ding{52}                               &\ding{52}                     &\ding{52}          \\ \hline
\end{tabular}
\end{center}
\end{table*}

\subsection{Overview of Provenance Graph-Based APT Audit}

\emph{Provenance Graph.} As shown in Fig.~1, a provenance graph \emph{G=\{N,\;E\}} is a directed graph enriched with chronological information, serving to capture and depict the interactions and causal relationships among diverse system entities including processes, files, {and} network connections. The graph \emph{G} is constructed through the collection of system logs from sources such as Windows ETW and Linux Auditd using probes (e.g., CamFlow) run on the OS \cite{5}. %which is usually stored in the format of json.
These logs provide the foundation for modeling large-scale system entities and their intricate interdependencies. The provenance graph then becomes a comprehensive representation of how these entities interact over time.
\begin{itemize}
    \item \emph{Entity.} It refers to the subject and object of system operations. In provenance graph auditing, as depicted in Fig.~1, the system entities mainly consist of three types: \emph{sockets} (or called network connections, represented as the parallelograms), files (represented as the rectangles), and processes (represented as the ellipses).
    \item \emph{Edge.} It refers to the causality dependency relationships between various entities, which primarily include \emph{read}, \emph{write}, \emph{execute}, and \emph{connect}. %(including operations such as send, receive, clone, etc.)
    For instance, in the provenance graph, the edge related to a file entity typically represents a read or write operation. In the case of a process entity, the edge usually indicates an execute operation; while for a socket entity, its edge typically represents a connect operation.
\end{itemize}

\subsection{State-of-the-Arts}

The pioneering work of SLEUTH \cite{10} introduced the provenance graph approach for real-time APT attack scenario reconstruction, by leveraging causal relationship tracking and provenance graph modeling. Besides, the attack process is reconstructed in \cite{10} through the construction and annotation of a lower-level event dependency graph. Subsequently, {novel algorithms for threat detection and heterogeneous graph construction were devised in NODOZE \cite{11}.} Poirot \cite{12} designed subgraph querying and matching algorithms {to address} the alignment challenge between APT attack primitives and provenance graphs. HOLMES \cite{13} innovatively merged the high-level scenario graph (HSG) with the ATT\&CK attack framework, thus resolving semantic alignment issues and effectively mitigating noise problems stemming from irrelevant sequences.
However, the efficiency shortcomings of the aforementioned approaches hinder the practical deployment of APT provenance graph auditing services.

The state-of-the-art literature on APT defense enhancements mainly focuses on three perspectives: reducing latency, countering highly covert APT behaviors, and causal relationship analysis. In terms of latency reduction, StreamSpot \cite{18} and UNICORN \cite{7} introduced a novel real-time runtime analysis framework for local hosts, which achieves attack detection without prior attack knowledge and demonstrates high accuracy with low false positive rates. Pertaining to defense against fileless attacks, ProvDetector \cite{8} introduced provenance graphs into concealed malicious attack detection and presented novel path algorithms to identify potential portions within provenance graphs, in order to establish recognition profiles for anomalous processes in each program. Through context of causal relationship analysis and natural language processing techniques, ATLAS \cite{2} proposed a sequence-based model based on audit logs, facilitating the end-to-end attack story generation. Additionally, DEPIMPACT \cite{15} extended ATLAS by introducing attack dependency subgraph weights, exploiting the similarity and closeness of attack sequences to achieve provenance graph compression and efficient audit.

Nevertheless, the above advanced approaches primarily target host-level APT detection, failing to account for network-level (i.e., the entire network consisting of multiple hosts) provenance auditing, thus lacking collaborative defense strategies among hosts. Furthermore, current APT defense strategies are susceptible to intelligent attacks such as APT evasion and adversarial subgraphs, resulting in a significant decline in the effectiveness of provenance graph detection.
{Table~I shows the comparisons of our work with existing state-of-the-arts.}

\begin{figure*}[!t]
\setlength{\abovecaptionskip}{-0.0cm}
\centering
  \fbox{\includegraphics[width=17cm]{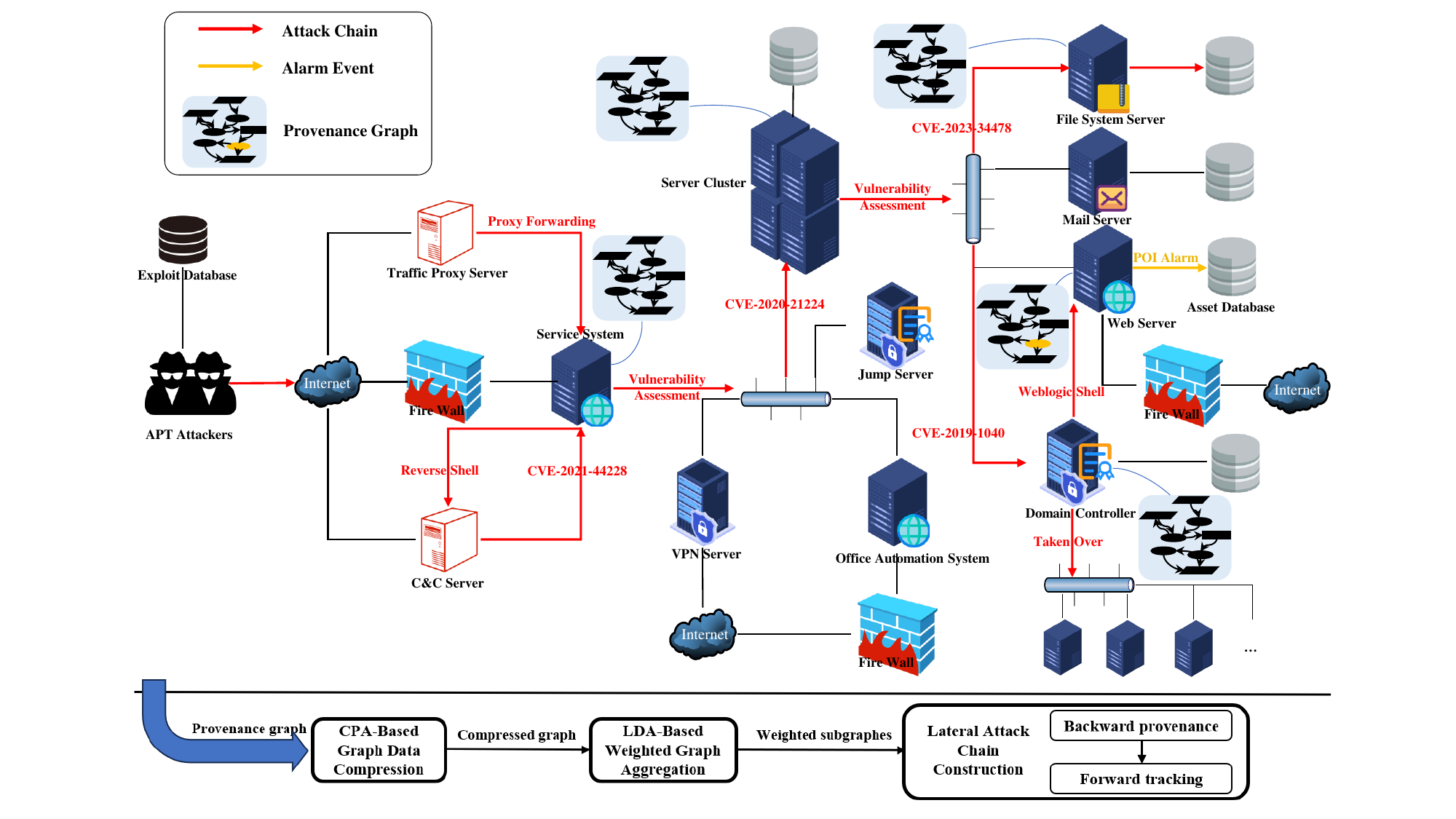}}
  \caption{{An Illustration of Network-Layer Distributed Provenance Graph Audit for Lateral Attack Chain Reconstruction.}}\label{fig:2}\vspace{-0.25cm}
\end{figure*}

\subsection{Challenges of Provenance Graph-Based APT Audit}

\begin{itemize}
    \item \emph{Low-Cost Lateral Attack Chain Reconstruction at the Network Level.} APT attacks are typically characterized by the high degree of stealth and prolonged persistence. A significant challenge is efficiently filtering relevant data from millions of provenance logs and establishing meaningful correlations to rapidly reconstruct APT attack chains. Current provenance graph audit schemes are confined to single-host operating systems, whereas real APT attacks exhibit a highly organized nature, often involving distributed and multi-point infiltrations.
    Relying solely on the auditing of a single host is inadequate to comprehensively reconstruct the complete attack event. Hence, it is imperative to devise a network-level collaborative provenance audit approach involving multiple hosts, while effectively compressing and  aggregating the extensive and multi-source provenance graphs. This is beneficial to the cost-effective reconstruction of APT lateral attack chains within complex and dynamic scenarios.

    \item \emph{Dynamic Detection of APT Evasion Behaviors with Temporal Correlations.} A team of APT attackers frequently employ various evasion strategies, such as interspersing numerous unrelated IPC sequences within attack primitives, to evade audit approaches based on provenance graphs. Consequently, the availability of APT detection services diminishes. However, existing provenance graph auditing approaches rarely account for APT evasion attacks, resulting in a demand for large-scale and {fine-grained} APT evasion behavior identification. Due to the massive and multi-source provenance graphs across diverse platforms, temporal correlations of {entities'} interactions in provenance graphs, and poisoning behaviors for targeted manipulation, it is challenging to design rapid stealthy evasion behavior detection mechanisms in such dynamic and uncertain environments.

    \item \emph{High-Robust and Self-Adaptive Adversarial Subgraphs Defense.} Existing APT defense strategies based on provenance graphs commonly rely on subgraph matching mechanisms, rendering them susceptible to adversarial attacks. {Adversarial attacks in provenance graph-based APT audit refer to a kind of sophisticated tactics that involve constructing adversarial provenance subgraphs \cite{9}. These attacks aim to evade subgraph matching-based detection mechanisms, while avoiding compromise of the attack primitives. As a result, they undermine the reliability of APT detection outcomes. A representative example of adversarial attacks in provenance graph-based APT audit using the DARPA TC dataset can be found in Fig.~1 in \cite{9}.}
        Adversarial attacks are scarcely considered in current works. As a result, there is a rising need for countering adversarial attacks. Given the concealed nature of adversarial subgraphs, the diversity of adversarial attack patterns, and the real-time and dynamically transmissible requirements of defense strategies, the design of robust and self-adaptive defense mechanisms against adversarial subgraphs is challenging.
\end{itemize}

\section{Solutions to Provenance Graph-Based APT Audit}
Aimed to address the challenges of lateral movement reconstruction, evasion behavior detection, and adversarial subgraph defense in current provenance graph based APT defense,
this section delves into the perspective of cost-effective and robust provenance graph based APT defense approaches, including network-level distributed provenance graph audit model (Sect.~\ref{solution1}), trust-oriented dynamic APT evasion behavior detection strategy (Sect.~\ref{solution2}), and HMM-based adversarial subgraph detection strategy (Sect.~\ref{solution3}).

\subsection{Network-Layer Distributed Provenance Graph Audit}\label{solution1}
In this subsection, we devise a distributed provenance graph audit model to efficiently reconstruct lateral attack chains from two perspectives: network-level global auditing and graph data compression. As shown in Fig. 2, it encompasses (i) a graph data compression module based on causality preserved aggregation (CPA) to address the issue of graph dependency explosion, (ii) a graph weight aggregation module based on linear discriminant analysis (LDA) to construct weighted provenance graphs, and (iii) a distributed APT lateral attack chain construction module using weighted provenance graphs.

\emph{1) CPA-Based Graph Data Compression:}
The CPA algorithm is utilized to effectively streamline the dependencies within the provenance graph involving extensive volume of data entities (e.g., IPC and file). Specifically, for two interconnected entity flows ($\rightarrow U\rightarrow V\rightarrow$) with a dependency relationship, the following three conditions are considered.
\begin{itemize}
  \item \emph{Forward ingress aggregation condition:} When the occurrence times of all ingress event edges into entity \emph{U} precede the event edge $U\rightarrow V$, the timestamp of the last ingress edge is designated as the global ingress time.
  \item \emph{Backward egress aggregation condition:} When the occurrence times of all egress event edges from entity \emph{V} follow the event edge $U\rightarrow V$, the timestamp of the initial egress edge is designated as the global egress time.
  \item \emph{Bi-egress aggregation condition:} For entity flows that meet both forward and backward aggregation conditions, the two entities are equivalently aggregated as the same one. {Besides, according to \cite{5}, daemon-related subgraphs within the provenance graph can form separate entities independently of other subgraphs, which can be removed to improve audit efficiency.}
\end{itemize}

\emph{2) LDA-Based Weighted Graph Aggregation:}
{This} model is used to trace PoI alert events by constructing weighted sub-provenance graphs. Three primary features, i.e., file size correlation, temporal relevance, and in-out degree ratio, are employed for extracting the entities in the provenance graph. Subsequently, the edges of the provenance graph are clustered through the multi-round K-means++ algorithm. Next, the LDA model is employed to compute the projection vectors that maximize the Fisher criterion for alarm-related edges versus non-alarm-related edges within the differentiated two groups of edges. {Then, the weight of each edge can be derived.}

\emph{3) Lateral Attack Chain Construction via Weighted Provenance Graphs:}
{Consider} the bidirectional interactivity in APT attack chains (i.e., the triggering of a PoI alert at the entry point evolves into a positive propagation toward linked sockets), \emph{socket} (for network connections) is the primary element.
{For a given PoI alert, this phase involves two successive steps:
\begin{itemize}
  \item \emph{Backward provenance:} Prioritize the weights of socket entities and select the highest-ranking one as the attack entry regarding the PoI alert event.
  \item \emph{Forward tracking:} Starting from the PoI alert event, compute and propagate the impact factor (IF) until subsequent entities in the provenance graph meet conditions 1, 2, and 3 simultaneously.
\end{itemize}
Here, the IF is inversely proportional to the magnitude of out-degrees. Then, we obtain the attack egress regarding the PoI alert event and halt the IF propagation. The IF is updated for the subsequent layer of incoming events, which serves as discriminative markers for lateral movement to help restore the corresponding APT lateral infiltration chain.}
{\emph{Condition 1} means that the IF of subsequent entities in the provenance graph surpasses the preset IF threshold.}
{\emph{Condition 2} indicates that the last entity in the provenance graph is a socket entity.}
{\emph{Condition 3} means that the last entity in the provenance graph differs from the socket entity obtained from the backward provenance.}

\begin{figure*}[!t]
\setlength{\abovecaptionskip}{-0.0cm}
\centering
\fbox{\includegraphics[width=13cm]{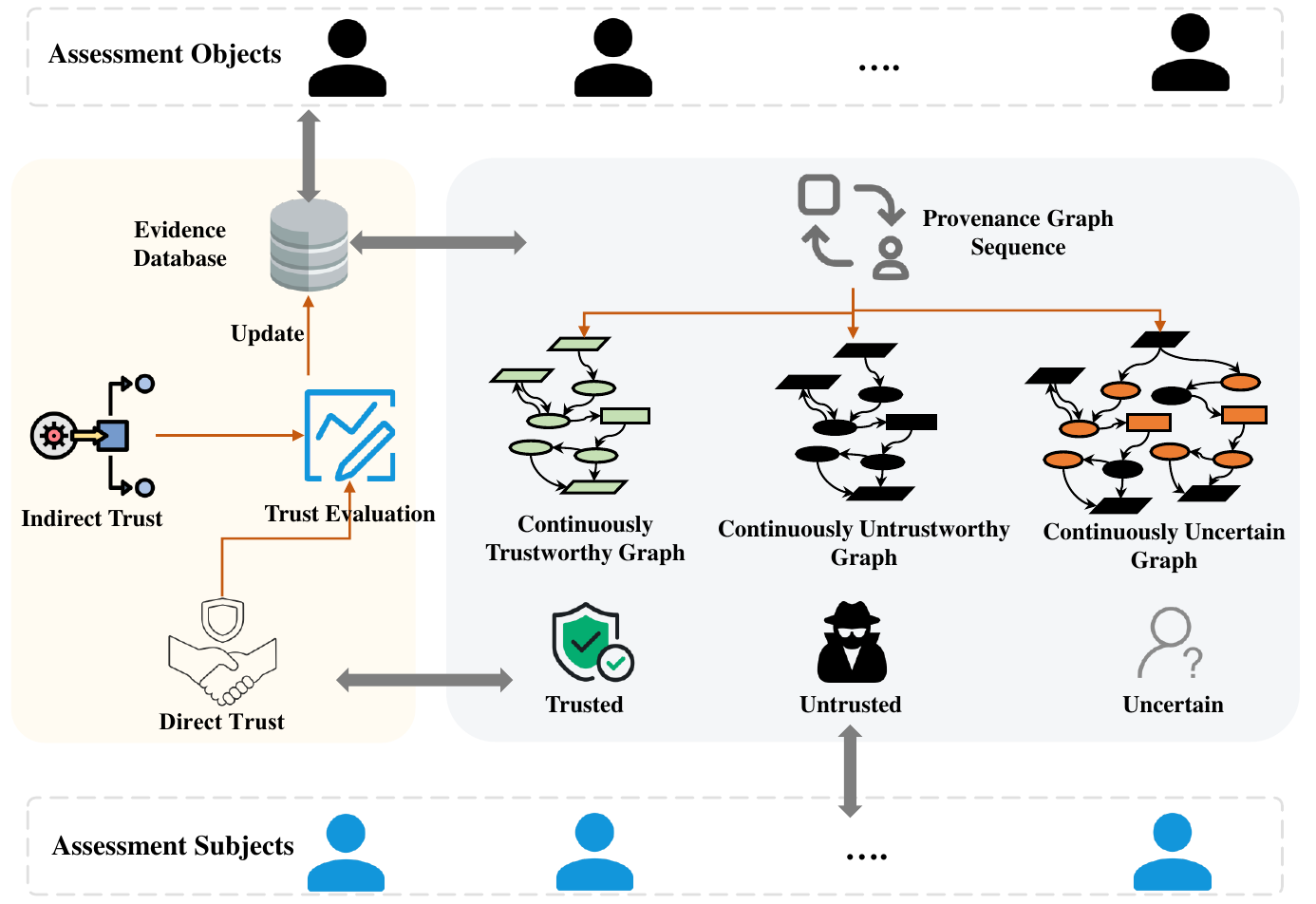}}
  \caption{{An Illustration of Trust-Oriented Dynamic APT Evasion Behavior Detection.}}\label{fig:3}
\end{figure*}

\subsection{Trust-Oriented Dynamic APT Evasion Behavior Detection}\label{solution2}
In this subsection, we devise a dynamic APT evasion behavior detection, which encompass (i) temporal correlation for attack-related substructure optimization in provenance graph, and (ii) dynamic trust assessment for suppressing behavioral sequences from untrusted entities.

\emph{1) Optimized Attack-Related Substructures of Provenance Graph:}
Adversaries can launch \emph{APT evasion attacks} by extending the completion time of their attack infiltration primitives and introducing irrelevant operations to saturate the payload entity flow with benign entities. Thereby, they can evade traditional pattern-matching based provenance graph detection {\cite{7},\cite{18}. To address this issue, a \emph{forgetting factor} for PoI alert events is introduced, which is associated with the penalty coefficient, the current time slot, and historical interactions. The penalty coefficient for an attacker represents the number of detected attack subgraphs within a specific time window (the length of which depends on the value of the forgetting factor).
For an attacker, if his penalty coefficient surpasses a predefined threshold, the causal dependencies of his distributed attack primitives can be temporarily correlated via the stack.
This allows the construction of provenance entity links related to the original attack behaviors, resulting in an optimized attack-related substructure within the original provenance graph. Furthermore, it helps reduce the impact of intentionally introduced benign entities by adversaries during trust evaluation.

\emph{2) Dynamic APT Evasion Behavior Analysis based on Trust Evaluation:}
As shown in Fig. 3, a defender (i.e., assessment subject) can obtain a sequence of optimized provenance graphs about an attacker (i.e., assessment object) from the evidence repository.
This sequence records the attacker's historical trustworthiness in chronological order, while each provenance graph in the sequence records the attacker's historical interactions with the victim host within a fixed time window.
Through sequence extraction methods, the sequence can be divided into three parts: subsequences of continuously trustworthy operations, continuously untrustworthy operations, and continuously uncertain operations. Then, we design a trust mechanism to distinguish an APT evasion attacker from an innocent user due to misoperations by evaluating the trustworthiness from both \emph{direct} and \emph{indirect} trust aspects. The direct trust is evaluated based on the Dempster-Shafer (D-S) evidence theory, considering the time span of continuously trustworthy/untrustworthy/uncertain operations and time decay effects.
{Generally, recent interactions in provenance graphs are more reflective of the current state of the trustee's behavior and intentions than older ones. Consequently,} we incorporate a \emph{time decay} factor into trust assessment, assigning greater weight to recent interactions and underscoring their influence on trust evaluation.
{Specifically, the exponential time decay function is employed to model the impact of interaction age on trust evaluation, taking into account both the decay rate and the time elapsed since the interaction time.}
Besides, we consider the \emph{rewarding and punishing effects} for continuous sequence, wherein users receive rewards for consistently providing trustworthy interactions and penalties for consistently engaging in malicious or uncertain behaviors.}
The indirect trust obtained from third-party recommendations can help enhance the accuracy of trust evaluation, especially when direct interactions are infrequent \cite{17}. Afterward, the latest trust evaluation results are stored in the evidence database.

\begin{figure*}[!t]
\setlength{\abovecaptionskip}{-0.0cm}
\centering
  \fbox{\includegraphics[width=13cm]{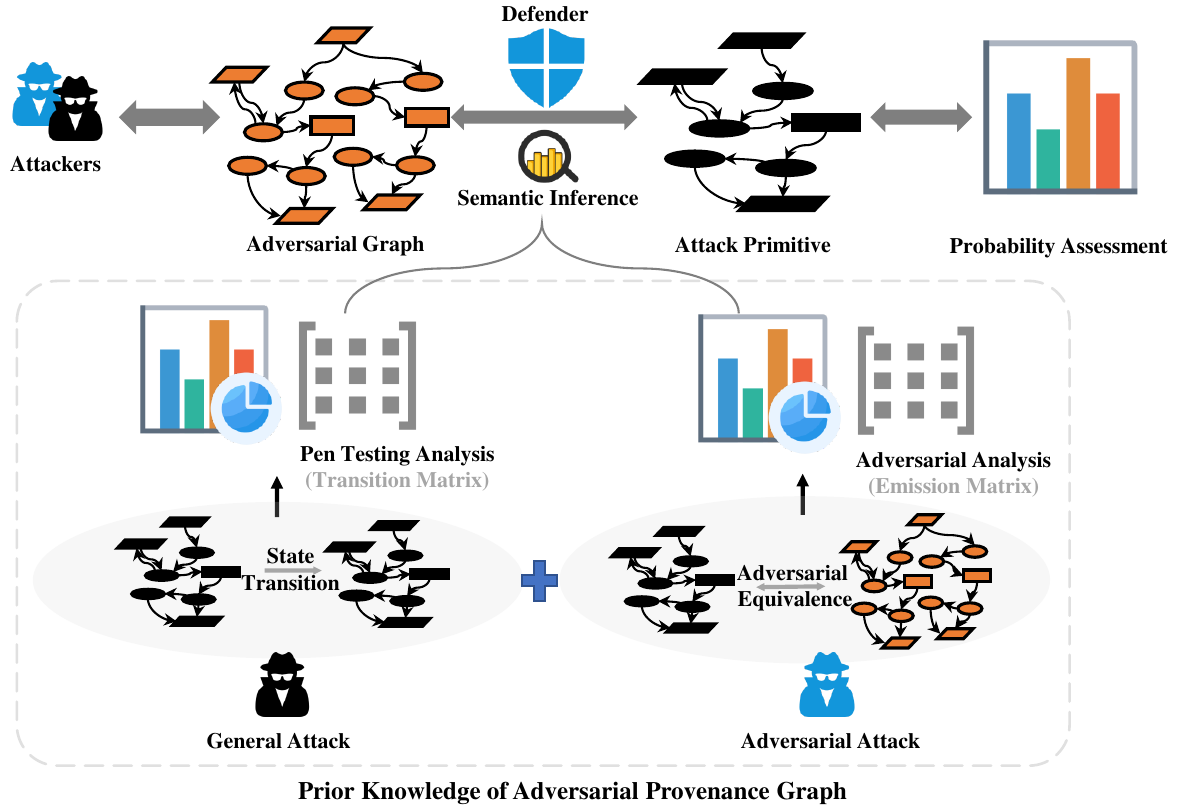}}
  \caption{{An Illustration of HMM-Based Adversarial Subgraphs Defense.}}\label{fig:4}
\end{figure*}

\subsection{HMM-Based Adversarial Sub-provenance Graph Defense}\label{solution3}
This subsection devises (i) a fast adversarial subgraph modeling method to explore adversaries' evasion principles during infiltration attacks, and (ii) a HMM-based self-evolving adversarial subgraph detection algorithm.

\emph{1) Fast Adversarial Subgraphs Modeling.} It comprises three steps.
\textit{Step 1: Test model construction based on subgraph matching.}
We train a general test {artificial intelligence (AI)} model for discriminating adversarial subgraphs, by optimizing the loss function, which is defined as one minuses the average number of successful attack subgraph matches for all subgraphs.
\textit{Step 2: Proof-of-concept framework design for adversarial subgraphs.} Initially, we utilize the subgraph deconstruction method \cite{12} to disassemble the subgraphs into individual substructures. These substructures are then summarized into an \emph{N}-dimensional vector using an encoding function. Subsequently, we employ a cosine distance-based discriminant function to determine if the subgraph is adversarial. This is achieved by comparing the cosine distance to a preset threshold.
\textit{Step 3: Adversarial subgraphs construction.} This step aims to create adversarial subgraphs without disrupting the original attack primitives. Initially, we select benign substructures to replace parts of the original graph's structure, with the objective of minimizing the cosine distance. Then, we update the cosine distance by applying the distance discriminant function to the modified subgraph. The above processes are repeated until the test model incorrectly classifies the subgraph as normal, resulting in the generation of an adversarial subgraph.

\emph{2) Robust Adversarial Subgraphs Detection Based on HMM.}
As depicted in the lower left of Fig.~4, we first construct a general attack subgraph using the ATT\&CK model\footnotemark[2] and the DARPA transparent computing dataset\footnotemark[3].\footnotetext[2]{\url{https://attack.mitre.org/}}\footnotetext[3]{\url{https://github.com/darpa-i2o/Transparent-Computing}}
Next, we count the attack directions (i.e., the potential entities to be linked next in the graph) within the APT to create a transfer matrix. Then, as shown in the lower right of Fig.~4, based on the adversarial equivalent graph obtained from our proposed fast modeling method, we count the adversarial transformation entities (i.e., the benign entities which are equivalent to the malicious entities) to derive the emission matrix. Finally, utilizing the obtained transfer matrix and emission matrix, we use the HMM Viterbi algorithm on the captured stream of provenance graphs to determine the most probable sequence of attack entities (i.e., those with the highest hit rate). When the hit rate surpasses a predefined threshold, the entity is identified as an adversarial subgraph.

\section{Implementation and Evaluation}

\subsection{Experimental Setup}
We implement an APT penetration test prototype with 15 servers to emulate a real {three-layer} enterprise internal network. {Each server is configured with 10 GB memory, 12th Gen Intel (R) Core (TM) i7-12700KF 3.60 GHz with 2-core, and Ubuntu 22.04.3 LTS. Besides, Camflow is employed to collect provenance graphs, and Neo4j serves as the database management system for provenance graphs.} Six types of attack scenarios are considered: \emph{buffer overflow}, \emph{domain controller hijacking}, \emph{living-off-the-land binaries (LOL bins)}, \emph{data leakage}, \emph{maintaining access}, and \emph{middleware exploitation}. These vulnerabilities are distributed across 15 servers, and each server is equipped with a lightweight provenance graph interface. The network-layer APT lateral movements, APT evasion attacks, and adversarial subgraph attacks are considered in our prototype.
{\begin{itemize}
  \item \emph{Efficiency.} We evaluate the system efficiency in terms of \emph{(i) the successfully restored lateral attack chains under various APT modes}, \emph{(ii) the reconstruction time of lateral attack chain reconstruction}, and \emph{(iii) trust evaluation for evasion behavior detection}.
  \item \emph{Robustness.} We evaluate the system robustness against adversarial subgraph attacks in terms of \emph{the recall rate {(also known as the true positive rate)} in the adversarial setting under various attack scenarios}. {Taking the LOL bins scenario as an example, its recall rate is the proportion of truly detected LoL bins threats out of the total attempts (i.e., the total number of truly/incorrectly detected LOL bins threats). The system robustness is measured by averaging recall rates across these scenarios, indicating the average defensive performance against adversarial attacks.}
\end{itemize}}

\begin{figure*}[!t]\vspace{-0.2cm}
\fbox{\centering
\subfigure[]{
\label{fig:5-1}
\includegraphics[width=6cm]{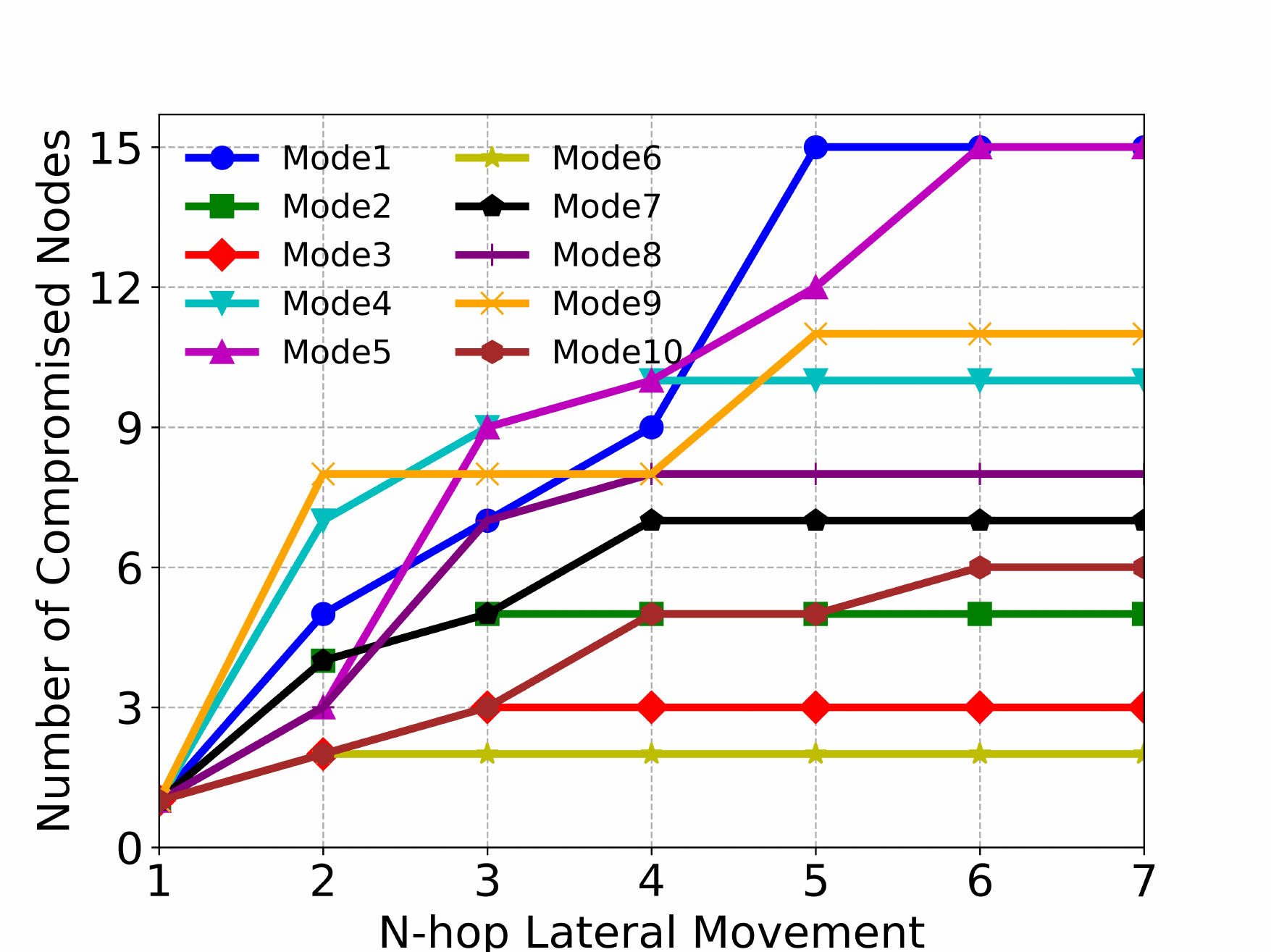}}
\subfigure[]{
\label{fig:5-2}
\includegraphics[width=6cm]{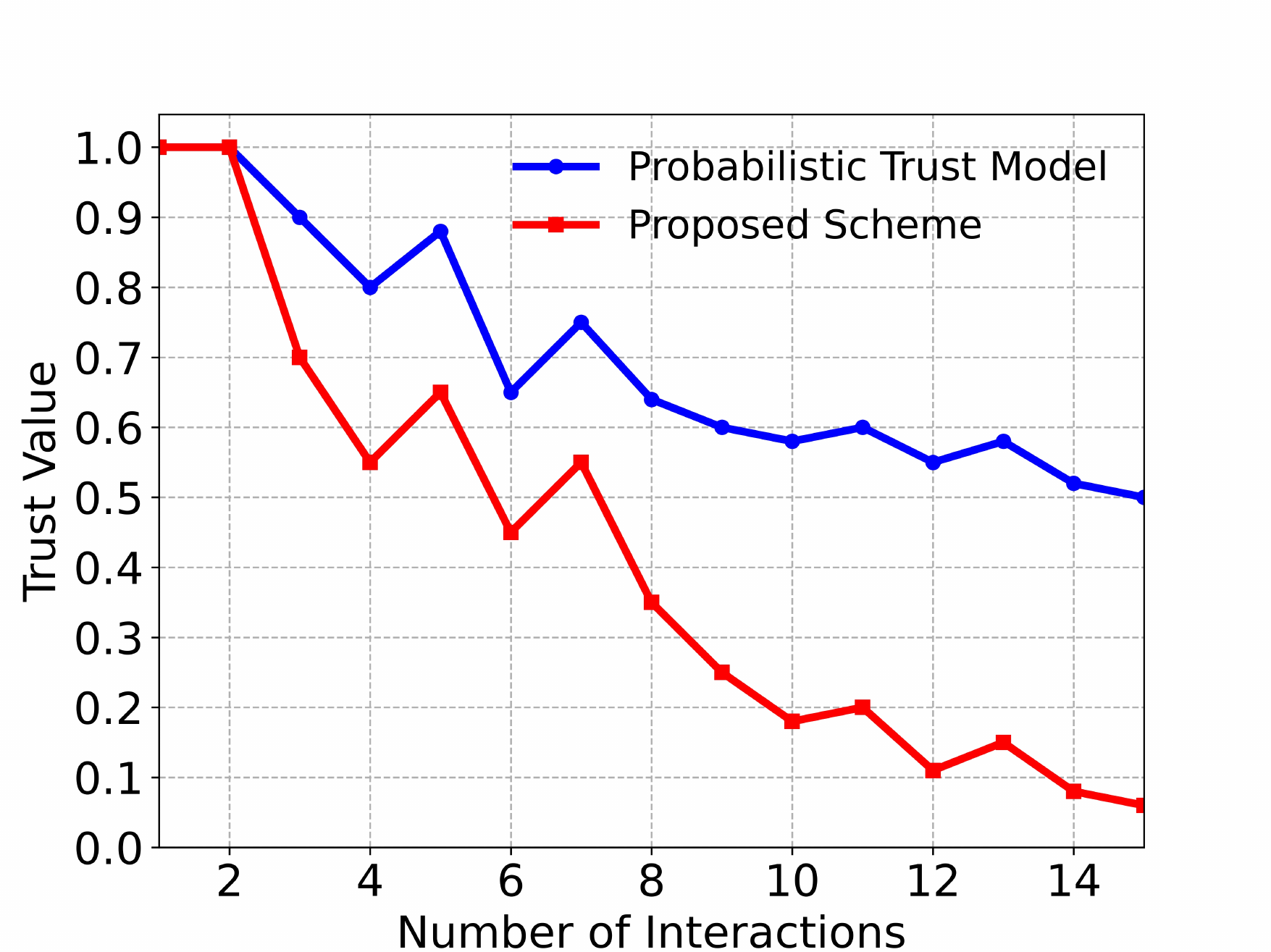}}
\subfigure[]{
\label{fig:5-3}
\includegraphics[width=6cm]{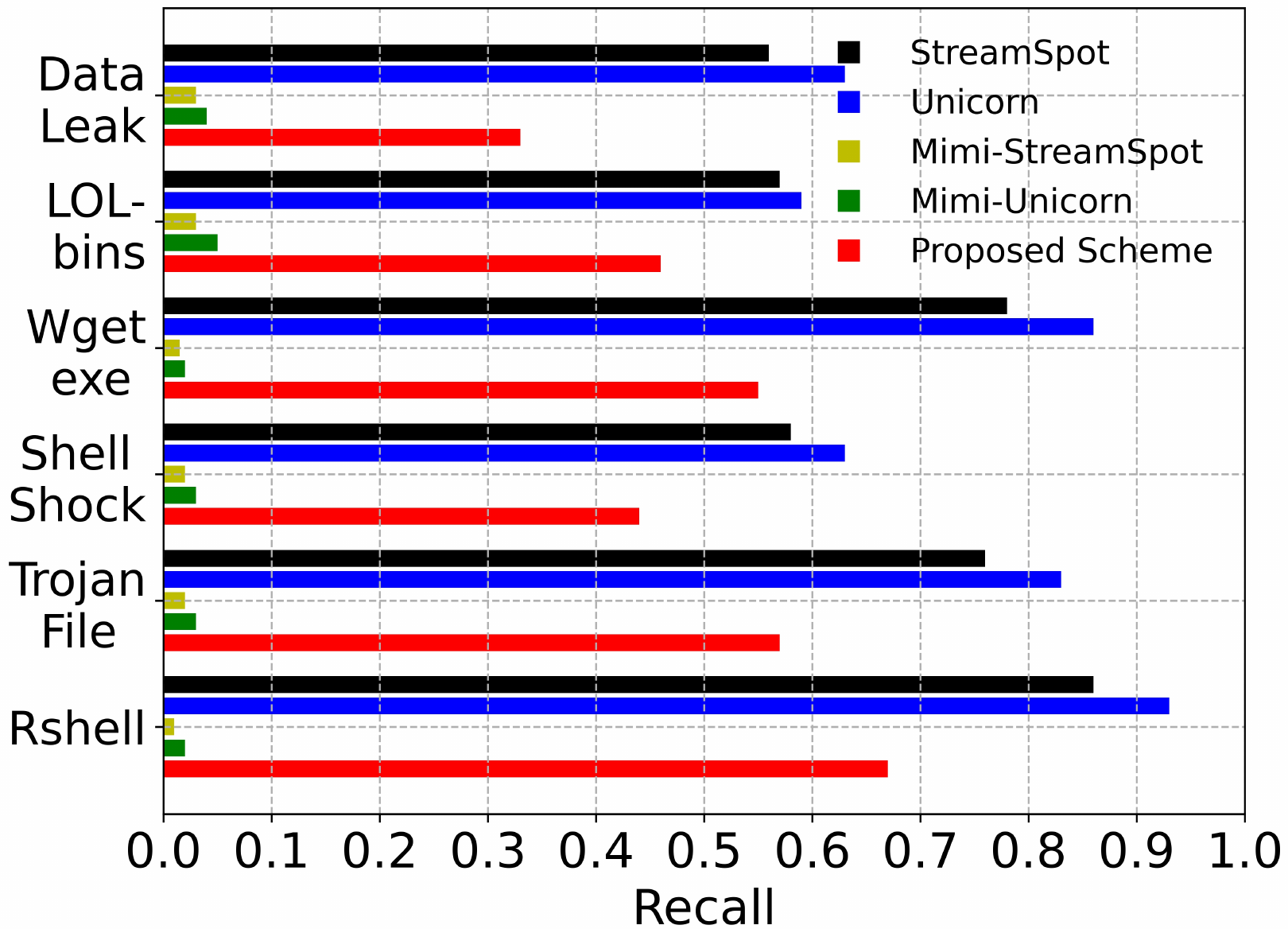}}}
\caption{a) Number of compromised nodes vs. \emph{N}-hop lateral movement under 10 modes of attack paths.
b) Trust value vs. number of interactions, compared with probabilistic trust model \cite{17}.
c) Successful detection rate w/wo adversarial attacks, compared with StreamSpot {\cite{18}}, Unicon {\cite{7}}, mimicry-StreamSpot, and mimicry-Unicon. \emph{Note:} As StreamSpot and Unicon are designed for the settings without adversarial attacks, to be fair and objective, we implement them under adversarial settings and call their modified versions as \emph{mimicry-StreamSpot} and \emph{mimicry-Unicon}, respectively.}
\label{fig:5}\vspace{-2mm}
\end{figure*}

\subsection{Experimental Results}
{In the experiments, we configure 10 APT attack modes {within each scenario}, each allowing a maximum of 6 lateral movements.} Fig.~5(a) illustrates the number of compromised nodes when the number of pivot servers (used for lateral movements) increases, {under 10 different APT modes. Notably, the proposed approach successfully restores lateral movement chains in all 10 APT modes. Besides,} different APT attack modes yield varying outcomes. For instance, in mode 1, the adversary executes a hijacking attack during the 4th round of lateral movement, successfully taking control of both the domain controller and domain users. In contrast, the adversary in mode 6 only succeeds in taking control of one server throughout the lateral movement attempts. {Additionally, the average time for APT attack chain reconstruction is under 3 minutes. In contrast, traditional approaches without CPA typically exceed 3 hours (and even several days in the worst case).}

{In trust evaluation, both the proposed scheme and the probabilistic trust scheme \cite{17} initially use historical interaction data to calculate the direct trust, and subsequently derive the integrative trust by aggregating all the recommendations from other nodes using D-S theory. However, \cite{17} neglects the varied impact of diverse trust evidence on the overall trust value and neglects that malicious nodes usually tend to consistently provide untrustworthy services.}
Fig.~5(b) shows the evolution of the trust value of the assessment object as the number of interactions increases under the APT evasion attack.
{As seen in Fig.~5(b), by integrating time decay effects and the continuous sequence with rewarding/punishing effects, the proposed approach outperforms the probabilistic trust model \cite{17} in achieving more accurate trust evaluation results, thereby detecting and punishing APT evasion behaviors more effectively.}

Fig. 5(c) illustrates the recall rates for different schemes with and without adversarial subgraphs under various attack scenarios. It can be seen that adversarial subgraphs significantly deteriorates the defensive effectiveness of conventional StreamSpot \cite{18} and Unicon \cite{7} schemes. Furthermore, {in the adversarial setting, the proposed scheme exhibits significant improvements in recall rates across the six typical attack types}, compared with the mimicry-StreamSpot and mimicry-Unicon approaches. Besides, the proposed scheme maintains a smaller performance gap compared to conventional StreamSpot and Unicon schemes without adversarial attacks. {As such, it validates the robustness of the proposed approach against adversarial subgraph attacks.}

\section{Future Directions}
This section explores the future directions that necessitate further research investigation in APT detection based on provenance graphs.
\subsection{Fusing Provenance Graphs and Knowledge Graphs for APT Detection}
Combining provenance graphs and knowledge graphs in APT detection is imperative to address semantic gaps and enhance threat provenance. Provenance graphs capture fine-grained system interactions, while knowledge graphs provide semantic context. This synergy offers comprehensive insights for accurate attack detection and attribution to achieve holistic and efficient APT detection. Effective fusion of heterogeneous data and knowledge representation remains the major challenge.

\subsection{Tamper-Resistant Provenance Graph Storage}
The integrity of the kernel-level provenance graph can be compromised by unauthorized modifications, posing risks to the reliability of audit trails. Cryptographic methods, such as digital signatures and secure hashing, offer a potential remedy against tampering threats. Immutable ledger technologies, such as blockchain, further bolster resistance to tampering by dispersing storage and enforcing consensus-based verification. Nevertheless, obstacles persist, encompassing efficient querying of encrypted data and managing access control in distributed environments. Developing tamper-resistant mechanisms is vital to upholding the trustworthiness of provenance-based APT audits.

\subsection{Collaborative and Privacy-Preserving Threat Intelligence Sharing}
APT threats often target multiple entities across sectors. Future research should focus on establishing collaborative frameworks for sharing CTI derived from provenance graph-based auditing. However, organizations may be hesitant to share sensitive data due to confidentiality and privacy issues, raising needs for privacy-preserving CTI aggregation and sharing without exposing sensitive information.
Other issues remain to be investigated include standardizing data formats and incentivizing collaboration.

\subsection{Integration with Cloud and Edge Environments}
Leveraging the integration of cloud and edge computing enhances APT detection services by enabling dynamic data correlation and analysis. Edge devices collect and preprocess local data for minimized latency{, and} cloud servers offer scalability and computational power for in-depth analysis and storage. This synergy optimizes APT detection, allowing real-time alerts at the edge and comprehensive analysis in the cloud. Research challenges include data synchronization, privacy preservation, and adaptation to spatiotemporal resource constraints in edges.

\section{Conclusion}
APT attacks have far-reaching consequences across various sectors including governments, critical
infrastructures and corporations, necessitating effective defense strategies. While existing provenance graph-based research sheds light on APT defense, the effectiveness of APT detection remains hindered by intricate lateral attack patterns, dynamic evasion strategies, and adaptive adversarial subgraphs. This study advocates for a novel approach for enhanced efficiency and robustness in existing provenance graph-based APT audit schemes, by devising a network-level distributed provenance graph audit model, a dynamic evasion behavior detection strategy, and a robust adversarial subgraph detection strategy. Via prototype implementation and experimental evaluations, the potential of the proposed system is validated to significantly enhance APT defense capabilities.
This work is anticipated to shed more light on ongoing exploration of comprehensive solutions against evolving APT threats in today's digital landscape.

\end{document}